\documentstyle{article}

\newcommand{\be}{\begin{equation}}
\newcommand{\ee}{\end{equation}}
\newcommand{\eqn}[1]{{\ref{#1}}}
\normalsize

\begin{document}

\bibliographystyle{unsrt}
\footskip 1.0cm
\thispagestyle{empty}
\setcounter{page}{0}
\begin{flushright}
IC/95/12\\
SISSA REF:17/95/EP\\
Napoli: DSF-T-3/95, INFN-NA-IV-3/95\\
February 1995\\
\end{flushright}
\vspace{10mm}

\centerline {\LARGE UV asymptotically free QED as a broken YM theory}
\vspace{5mm}
\centerline {\LARGE in the unitary gauge}
\vspace*{15mm}
\centerline {\large G. Bimonte$^{a,c}$, R. Iengo$^{b,d}$}
\vspace{5mm}
\centerline{ \small and }
\vspace*{5mm}
\centerline{ \large G. Lozano$^a$ \footnote{
E-mail addresses:
Bimonte@ictp.trieste.it~~,~~iengo@sissa.it~~,~~Lozano@ictp.trieste.it}}
\vspace*{5mm}
\centerline {\it $^a$ International Centre for Theoretical Physics, P.O.BOX
586}
\centerline {\it I-34100 Trieste, ITALY}

\vspace*{5mm}
\centerline {\it $^b$ International School for Advanced Studies - SISSA}
\centerline{ I-34014 Trieste, ITALY}

\vspace*{5mm}
\centerline {\it $^c$ INFN, Sezione di Napoli, Napoli, ITALY}

\vspace*{5mm}
\centerline {\it $^d$ INFN, Sezione di Trieste, Trieste, ITALY}

\vspace*{25mm}
\normalsize
\centerline {\bf Abstract}
\vspace*{5mm}
{\large We compute the $\beta$-function of a YM theory, broken to $U(1)$,
by evaluating the coupling constant renormalization in
the broken phase. We perform the calculation  in the unitary gauge where
only physical particles
appear and the theory looks like a version of QED containing
massive charged spin 1 particles. We consider an on-shell scattering
process and after verifying that the non-renormalizable divergences which
appear in the Green's functions cancel in the expression of the amplitude,
we show  that
the coupling constant renormalization is entirely due to the photon
self-energy as in QED. However we get the expected asymptotic freedom
and the physical charge decreases logarithmically as a
function of the symmetry breaking scale.}

\newpage

\baselineskip=24pt
\setcounter{page}{1}

Here we present a computation of the renormalization group
$\beta$-function for a QED-like theory, i.e. an unbroken $U(1)$ gauge
theory, in which there are charged massive vector bosons.
Renormalizability  requires it to be the broken phase of a
non-abelian theory (which here for definiteness we consider to be an $SU(2)$
gauge theory with a Higgs in the adjoint representation, i.e. the well
known Georgi-Glashow model \cite{geo}).  Since the $\beta$ function is
related
to the high energy properties of the model, it is expected to coincide
with the
asymptotically free one computed in the symmetric phase \cite{gro}.
Here we perform the calculation from the point of view of the low energy
$U(1)$ phase, choosing  the unitary gauge  and thus including in the loops
only physical and (except for the
$U(1)$ photon) massive particles. As far as we know, this computation has
not been done before. From one side, it allows to express the $\beta$
function in terms of the contributions of the physical degrees of freedom
of the broken phase while, from the other side, it illustrates the subtle
mechanism by which an
abelian theory (the unbroken $U(1)$) with massive charged spin-1 bosons
manages to be asymptotically free. Recently, this property has been shown
to play a key role in understanding the duality properties of the low
energy phase of the $N=2$ SY theory (see \cite{sei}). There, what is
relevant is the dependence of the effective coupling constant on the
expectation value of the Higgs field, proportional to the charged spin-1
particle mass. We will see that this dependence comes out rather
directly in our computation. It is usually stated that the unitary
gauge represents an unconvenient choice for performing perturbative
calculations because of the bad high energy behavior of the massive
gauge particle propagator which causes the occurrence of non
renormalizable divergences in the computation of Green's functions. We will
show here that this fact does not represent any technical difficulty, but
on the contrary, the computation of the $\beta$ function turns out to be
remarkably simple.  Actually, our understanding of this point
has greatly benefitted from a very recent paper by J. Papavassiliou and A.
Sirlin \cite{sir},(see also \cite{deg})
which indicates what to expect for the cancellation mechanism of
non-renormalizable divergences  that is known to hold for the
computation of gauge invariant on-shell quantities \cite{ZJ}.
We make use of
this point of view by defining the physical coupling constant as the
residue of the pole at zero momentum transfer of the on-shell scattering
amplitude
for two charged fermions. After verifying the expected cancellation of
non renormalizable divergences (for which we
make use of available results for the computation of the various
graphs involved \cite{bar}), we show that
the $\beta$ function can be extracted just from the photon self-energy.
In particular, its asymptotically free part comes from a single
graph, namely the photon self-energy due to the charged massive
vector boson.
If we compare the
result with the standard $\beta$ function computation in the unbroken
phase, we see that this graph alone reproduces the sum
of the contributions to the $\beta$-function of the vector bosons and the
Higgs scalars.

It is perhaps worthwhile to notice that the same type of procedure,
namely considering scattering amplitudes to extract the $\beta$-function,
is also the starting point of string theory computations, see \cite{min},
since there, in principle, only on-shell $S$-matrix elements are defined.

To be specific we consider a $SU(2)$ theory broken to $U(1)$ by a Higgs
in the adjoint representation and containing two families of Dirac
fermions in the fundamental representation. The corresponding Lagrangian
is:
$$
{\cal L}=-\frac{1}{4}G^a_{\mu\nu}G^{a\mu\nu}+\frac{1}{2}
D_{\mu}\Phi^a D^{\mu}\Phi^a-\frac{\lambda}{4}(\Phi^a\Phi^a-v^2)^2+
$$
\be
+\overline{\psi}_f(i \not \!\! D -m_f)\psi_f~,\label{lag}
\ee
where
\be
G^a_{\mu\nu}=
\partial_{\mu}A_{\nu}^a -\partial_{\nu}A_{\mu}^a
-g\epsilon^{abc}
A_{\mu}^b A_{\nu}^c~~,
\ee
\be
D_{\mu}\Phi^a=\partial_{\mu}\Phi^a-g\epsilon^{abc}A_{\mu}^b\Phi^c~,
\ee
\be
D_{\mu}\psi_f=\left(\partial_{\mu}+i\frac{g}{2}\tau^a A_{\mu}^a
\right)\psi_f~,
\ee
with $\tau^a$ the Pauli matrices, $(a=1,2,3)$ and $f=1,2$ a family
index.
After the breaking $SU(2) \Longrightarrow U(1)$, we have in the unitary
gauge:
\be
\Phi^1=\Phi^2=0~,~~~~~~\Phi^3(x)=v+\phi(x)~.
\ee
Defining $W_{\mu}^{\pm}=\frac{1}{\sqrt 2}(A^1_{\mu}\mp i A^2_{\mu})$
to be the charged vector boson fields and $A_{\mu}=A^3_{\mu}$ to be the
photon field,
 we can rewrite the Lagrangian
in the form of a $U(1)$ theory:
\newpage
$$
{\cal L}=
-\frac{1}{4}(\partial_{\mu}A_{\nu} -\partial_{\nu}A_{\mu})^2
-\frac{1}{2}|\partial_{\mu}W_{\nu}^+ -\partial_{\nu}W_{\mu}^+|^2-
M_W^2 W^+_{\mu}W^{-\mu} +\overline{\psi}_f(i \not \! \partial - m_f)\psi_f~+
$$
$$
+ig(\partial^{\mu}A^{\nu} -\partial^{\nu}A^{\mu})W^+_{\nu}W^-_{\mu} +
(igA^{\mu}(\partial_{\mu}W_{\nu}^+ -\partial_{\nu}W_{\mu}^+)W^{-\nu}+
h.c)+
g^2(A^{\mu}A^{\nu}W^+_{\mu}W^-_{\nu} -
A_{\nu}A^{\nu}W^+_{\mu}W^{-\mu})+
$$
\be
-\frac{g}{2}A_{\mu}\overline{\psi}_f\gamma^{\mu}\tau^3\psi_f -
\frac{g}{\sqrt 2} (W_{\mu}^+\overline{\psi}_f\gamma^{\mu}\tau_+
\psi_f+h.c.)+\cdots  \label{lag2}
\ee
Here $\tau_+=1/2(\tau^1+i\tau^2)$ and $M_W = g v$ is the mass acquired by
the W bosons after symmetry breaking. In eq.\eqn{lag2}, we have omitted the
terms of the
lagrangian which do not contribute, at one loop, to the process we are
interested in. Clearly, in the unitary gauge all the fields appearing in
the Lagrangian correspond to physical particles. One can think of the
Lagrangian, eq.\eqn{lag2}, as describing a particular version of QED
including a massive charged vector boson.

As we said above, we will consider a definite physical process, the
on-shell elastic scattering of a positively charged fermion of the first
family off a negatively charged fermion of the second family.
Let us notice that the scalar field is not coupled to the fermions and
that it is not involved in any of the diagrams that contribute to
this process
at the 1-loop order. In addition, as we consider the elastic scattering
of two distinguishable fermions, there are neither annihilation nor exchange
channels.

Figure 1 displays the graphs involving $W$ bosons which contribute to the
process at the
order $g^4$ (Of course, to get the full amplitude, one has to add also
the ordinary diagrams of
spinorial electrodynamics, which are not shown for brevity). These
graphs can be grouped in four classes:
\begin{enumerate}
\item the graphs $P_1,P_2$ represent the $W$ contribution to the photon
self-energy. $P_2$ is a tadpole like graph and only $P_1$, where a
$W^{\pm}$ pair is created and then annihilated, gives rise to the photon
wave function renormalization.
\item the graphs $E_1,\cdots , E_4$ represent the contribution of the $W$
to the fermion wave function renormalization.
\item the graphs $V_1,\cdots ,V_4$ represent the contribution of the $W$
to the radiative correction of the photon-fermion vertex.
\item the graph $B$ is a box diagram, where a $W^+$ and a $W^-$ are
exchanged in the momentum transfer channel.
\end{enumerate}

We shall use this scattering process in order to define the physical
coupling constant $g_{ph}$. Calling ${\cal M}$ the scattering amplitude,
we define $g_{ph}$ by means of the residue at zero momentum transfer
$q^2=0$ :
\be
g^2_{ph}=i \lim_{q^2\rightarrow 0}[q^2 {\cal M}]~\cdot 4
\frac{m_1m_2}{(p_{1 \mu} p_2^{\mu})}~~,\label{geff}
\ee
where $p_{1,2}$ are the four momenta of the incoming particles.

Now, the important point to notice is that of all the diagrams of Fig1,
the only one which contributes to the pole is $P_1$.
This is so because at $q^2 \rightarrow 0$
the pole part of the diagrams $V_1,\cdots , V_4$ cancels against the
contribution of the diagrams $E_1,\cdots , E_4$. This is a consequence of
the $U(1)$ Ward identity, which we have explicitly checked to hold on the
diagrams involving the $W$. As for the box diagram $B$ it does not have
any pole for $q^2\rightarrow 0$. Concerning the diagram $P_2$, it
corresponds to a double pole, which, by gauge invarince, cancels against
the double pole part of $P_1$.

The sum of the diagrams $P_1$ and $P_2$ gives the contribution:
\be
P_1+P_2=J_1^{\mu}\left(-i\frac{g_{\mu\rho}}{q^2}\right)
i\Pi^{\rho\sigma}
\left(-i\frac{g_{\sigma\nu}}{q^2}\right)J_2^{\nu}~~,\label{p1p2}
\ee
where
$$
J_1^{\mu}=-i\frac{g}{2}\overline{u}_1(p_1+q)\gamma^{\mu}u_1(p_1)~~
,~
J_2^{\nu}=i\frac{g}{2}\overline{u}_2(p_2-q)\gamma^{\nu}u_2(p_2)~~,
$$
are the fermion currents, and the vacuum polarization tensor is:
\be
\Pi_{\rho\sigma}=(-g_{\rho \sigma}q^2+q_{\rho}q_{\sigma})F(q^2)~.\label{vapo}
\ee
We thus have, using current conservation:
\be
P_1+P_2=i\frac{F(q^2)}{q^2}J_{1\mu}J_2^{\mu}~.
\ee
The Feynman graph $P_1$ corresponds to the following contribution to
$\Pi_{\rho \sigma}$:
\be
i~\Pi_{\rho \sigma}^{(P_1)}=(ig \mu^{\epsilon})^2~\int \frac{d^D k}{(2
\pi)^D}
V_{\rho \alpha \beta}(q,k,k+q)~iD^{\alpha \delta}(k)~iD^{\beta \gamma}(k+q)~
V_{\sigma \gamma \delta}(-q,k+q,k)~.\label{diag}
\ee
Here
\be
i D_{\mu \nu}(k)
=i\frac{-g_{\mu\nu}+\frac{k_{\mu}k_{\nu}}{M_W^2}}{k^2-M_W^2}~.
\label{pro}
\ee
is the $W$-propagator in the unitary gauge and
\be
V_{\lambda \mu \nu}(k_1,k_2,k_3)=g_{\lambda \mu}(k_1-k_2)_{\nu}+g_{\mu \nu}
(k_2+k_3)_{\lambda}-g_{\nu \lambda}(k_3+k_1)_{\mu}~.\label{ver}
\ee
We make the computation in the dimensional regularization scheme, where
$D=4-2\epsilon$ and $g \rightarrow g \mu^{\epsilon}$, with $\mu$ the
regularization scale.

Upon adding the contribution to $\Pi_{\rho \sigma}$ from the graph $P_2$,
which is $q$-independent and cancels a similar term from $P_1$, we get
the $U(1)$ gauge invariant expression eq.\eqn{vapo}, with
\be
F(q^2)=\frac{g^2}{16 \pi^2}\left[ -7+
\frac{7}{6}\frac{q^2}{M_W^2}+
\frac{1}{12}\frac{q^4}{M_W^4}\right]\left(\frac{1}{\epsilon}-\log\frac{M^2_W}
{4 \pi \mu^2}\right)+\frac{g^2}{16 \pi^2}G(q^2)~~.\label{F}
\ee
$G(q^2)$ is finite in the limit $\epsilon
\rightarrow 0$ with value at $q^2=0$:
$$
G(0)=\frac{2}{3}+7\gamma~~.
$$
$\gamma$ being the Euler constant.

The above result  eq.\eqn{F} was already available in the literature
\cite{bar} and
we have checked it by an independent computation.

We point out that the expression for $F$ contains non-renormalizable
divergent terms, namely those of order
$q^2/M_W^2$ and $q^4/M_W^4$.  The presence of such terms in the
expression of the Green's functions is a peculiarity of the unitary gauge
and it is in this sense that it is usually referred to as a non
renormalizable gauge. These terms originate from the bad ultraviolet
behavior of the massive vector boson propagator (eq.\eqn{pro}).
Those nonrenormalizable divergences cancel from the
on-shell scattering amplitude \cite{ZJ,sir, deg}. In fact, other graphs of
Fig.1 have also divergences higher than usual.
The vertex parts of the diagrams $V_1,\cdots,V_4$ have divergent terms
proportional to $q^2/M_W^2$ and $q^4/M_W^4$ (to be precise, $V_1$ and
$V_3$ turn out to be finite), while the box diagram $B$
has a divergent term proportional to $1/M^2_W$ and another proportional to
$q^2/M^2_W$. By keeping into account the $1/q^2$ of the photon propagator
and adapting to our case the available expressions for those divergent terms
\cite{bar},
one can check that the nonrenormalizable divergences cancel from the
amplitude.

Thus, the only divergent term that remains in the amplitude is the one
occurring in $F(q^2=0)$ which is removed by the photon wave function
renormalization constant $Z$, as it happens in ordinary QED. Indeed
eq.\eqn{geff} gives to one-loop order:
\be
i \lim_{q^2\rightarrow 0}[q^2 {\cal M}]~\cdot 4
\frac{m_1m_2}{(p_{1 \mu} p_2^{\mu})}~=g^2(Z^{-1}-F(0)-F_{\it
QED}(0))~,\label{lim}
\ee
where $F_{\it QED}(q^2)$ is the standard contribution of the fermions in
spinorial QED. This allows us
to immediately read off this renormalization constant in the minimal
subtraction scheme:
\be
Z=1+7
\frac{g^2}{16 \pi^2}\frac{1}{\epsilon}+\delta Z_{\it QED}~,\label{Z}
\ee
where $\delta Z_{\it QED}$, the contribution of the two families of fermions
doublets, reads:
\be
\delta Z_{\it QED}=-\frac{4}{3} \frac{g^2}{16 \pi^2}\frac{1}{\epsilon}~.
\ee
By defining the renormalized coupling constant in terms of the bare one
$g_0$ as
\be
g=g_0 \mu^{-\epsilon}Z^{1/2}
\ee
(we have made use of the QED Ward identity) we compute the $\beta$ function
\be
\beta(g)=-\frac{g^3}{16 \pi^2}\left(7-\frac{4}{3}\right)~~.\label{beta}
\ee
The same result is obtained for the unbroken $SU(2)$ gauge theory, with a
scalar multiplet in the adjoint representation and two fermion families
in the fundamental representation. The number $7$ in eq.\eqn{beta}, which
represents the massive vector contribution precisely accounts for the
combined contribution of the $SU(2)$ massless gauge fields and Higgs scalars
of the unbroken phase: $-7=-22/3+1/3$. It
is amusing to note that while in the unbroken $SU(2)$ theory one gets the
$\beta$-function from the computation of several diagrams, including in
general both vertex and self-energy parts, in the $U(1)$ phase the same
result comes just from the evaluation of the photon self-energy.

It may be useful to stress that the term $k_{\mu}k_{\nu}/M_W^2$ in the
propagator eq.\eqn{pro}, while is responsible of the nonrenormalizable
divergences in eq.\eqn{F}, is also essential to obtain the correct result
for $q^2 \rightarrow 0$.

We may also further remark that it is the lowest, i.e. renormalizable,
divergence which contributes the negative term $-7/ \epsilon$ in $F(q^2)$
eq.\eqn{F}, while the highest non-renormalizable one has a positive sign.
In usual QED, without charged massive vectors, the highest divergence in
$F_{\it QED}(q^2)$
is the renormalizable one and it has a positive sign,
giving a term in the $\beta$-function which is opposite to UV asymptotic
freedom.

Upon inserting eq.\eqn{F} and the known expression for $F_{\it QED}(q^2)$
into eq.\eqn{lim} we get:
\be
g^2_{ph}=g^2\left(1-\frac{g^2}{16 \pi^2}
\left[7 \log \frac{M^2_W}{\tilde \mu^2}
+ \frac{2}{3}
-\frac{2}{3}\log \frac{m^2_1}{\tilde \mu^2}
-\frac{2}{3}\log \frac{m^2_2}{\tilde \mu^2}\right]
\right)~.\label{gef2}
\ee
where $\tilde \mu^2=4 \pi e^{-\gamma} \mu^2$. Here $g^2$ has to be
considered as a reference value, independent of the masses of the
particles. By summing over the iterations of the self-energy corrections,
eq.\eqn{gef2} can be recast in the following convenient form
\be
\frac{1}{g^2_{ph}}=\frac{1}{g^2}+\frac{1}{16 \pi^2}
\left[7 \log \frac{M^2_W}{\tilde \mu^2}
+ \frac{2}{3}
-\frac{2}{3}\log \frac{m^2_1}{\tilde \mu^2}
-\frac{2}{3}\log \frac{m^2_2}{\tilde \mu^2}\right]~.\label{gef3}
\ee
This equation exhibits the mass dependence, characteristic of the
effective coupling of a broken non-abelian theory, see for instance
\cite{bin}.

As a final remark, in the $N=2$ supersymmetric $SU(2)$ Y.M. theory,
broken to $U(1)$, one has to evaluate the r.h.s. of eq.\eqn{gef3} by
taking into account, in the place of our fermion families, the
supersymmetric partners of the $W$. One thus gets \cite{sei}:
\be
\frac{1}{g^2_{ph}}=\frac{1}{g^2}+\frac{1}{4 \pi^2}\log \frac{M^2_W}
{\tilde \mu^2}~.
\ee

\newpage

{\bf Acknowledgements}

We would like to thank Prof.  Salam, the International Atomic Energy
Agency and UNESCO for hospitality at the International Centre
for Theoretical Physics.

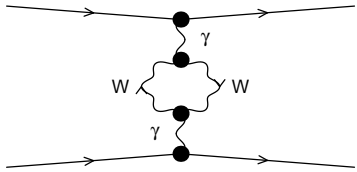
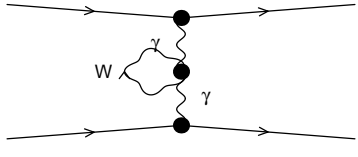
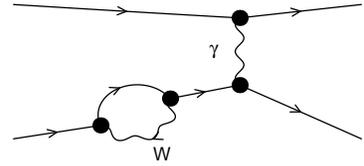
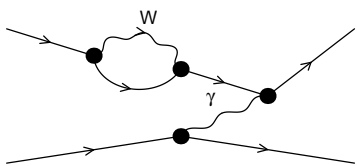
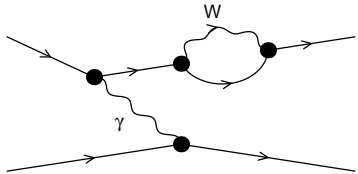
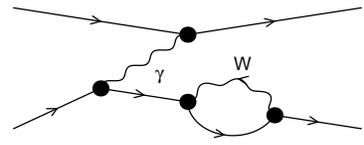
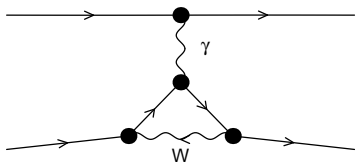
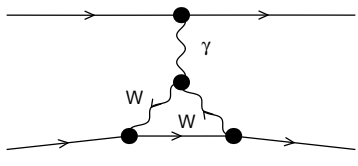
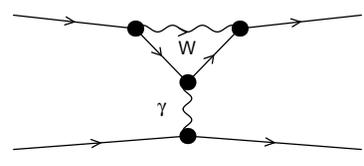
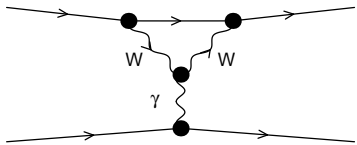
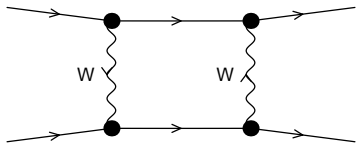
\end{document}